\documentstyle[12pt]{article}
\begin{document}
\def\a{\alpha}
\def\b{\beta}
\def\ch{\chi}
\def\d{\delta}
\def\e{\epsilon}
\def\E{{\cal E}}
\def\f{\phi}
\def\g{\gamma}
\def\h{\eta}
\def\et{\tilde{\eta}}
\def\i{\iota}
\def\j{\psi}
\def\k{\kappa}
\def\l{\lambda}
\def\m{\mu}
\def\n{\nu}
\def\o{\omega}
\def\p{\pi}
\def\q{\theta}
\def\r{\rho}
\def\s{\sigma}
\def\t{\tau}
\def\u{\upsilon}
\def\x{\xi}
\def\z{\zeta}
\def\D{\Delta}
\def\F{\Phi}
\def\G{\Gamma}
\def\J{\Psi}
\def\L{\Lambda}
\def\O{\Omega}
\def\P{\Pi}
\def\S{\Sigma}
\def\U{\Upsilon}
\def\X{\Xi}
\def\T{\Theta}
\def\vf{\varphi}
\def\ve{\varepsilon}
\def\cC{{\cal P}}
\def\cD{{\cal Q}}

\def\pp {\partial }
\def\pb {\bar{\partial }}
\def\zb {\bar{z}}
\def\ET {\tilde{E }}
\def\be{\begin{equation}}
\def\ee{\end{equation}}
\def\ben{\begin{eqnarray}}
\def\een{\end{eqnarray}}
\def\cos{\mbox{cos}}
\def\sin{\mbox{sin}}
\def\cot{\mbox{cot}}
\def\Tb{\bar{T}}
\hsize=16truecm
\addtolength{\topmargin}{-0.6in}
\addtolength{\textheight}{1.5in}
\vsize=26truecm
\hoffset=-.6in
\baselineskip=7 mm
\rightline{\today} \vskip2cm
\centerline{\Large\bf Vortex Strings and Nonabelian sine-Gordon Theories}
\begin{center}
\vskip 1cm
{ Q-Han Park\footnote{e-mail:
qpark@nms.kyunghee.ac.kr} and H.J. Shin\footnote{e-mail:
hjshin@nms.kyunghee.ac.kr}}
\vskip3mm
{Department of Physics and Research Institute of Basic
Science \\ }
{Kyung Hee University, Seoul 130-701,  Korea}
\vskip 2cm
{\bf ABSTRACT}
\end{center}
\vskip 5mm

We generalize the Lund-Regge model for vortex string dynamics in 4-dimensional
Minkowski space to the arbitrary n-dimensional case. The n-dimensional
vortex equation is identified with a nonabelian sine-Gordon equation and
its integrability is proven by finding the associated linear
equations of the inverse scattering. An explicit expression of
vortex coordinates in terms of the variables of the nonabelian sine-Gordon
system is derived. In particular, we obtain the n-dimensional vortex soliton
solution of the Hasimoto-type from the one soliton
solution of the nonabelian sine-Gordon equation.
\newpage
The relativistic motion of vortex strings in a superfluid was
first modeled by Lund and Regge in 1976 \cite{lund1}. Among many vortex
models, their model is distinguished in that it is an exactly integrable
model and it becomes the Nambu string model in the no-coupling
limit. In the context of string theory, the nonvanishing coupling term also received
an interpretation as describing the interaction of a string with background
antisymmetric tensor fields \cite{zee}.
Lund and Regge has proven the integrability of the model by recognizing the
vortex equation as the integrability equation of Gauss and
Codazzi\footnote{Many integrable equations arise from the study of the surface
embedding problem in differential geometry which provides a clear geometrical
meaning to integrable equations. For the modern formulation of surface embedding
problem, see for example in \cite{Bob}. } \cite{lund1,lund2}.
They have identified the vortex equation with the complex
sine-Gordon equation \cite{Pol} and found the associated linear equations
of the inverse scattering.  Since then, the complex sine-Gordon
theory has been studied intensively \cite{Get}-\cite{Park2}, with
applications to nonlinear optics \cite{Park3}. Extensions to more general
cases of the nonabelian sine-Gordon theories have been also made by associating
them with symmetric spaces \cite{bps} and their properties were investigated in
detail \cite{Park4,Hol}.

However, the vortex model by Lund and Regge is defined only in the 3+1-dimensional
Minkowski space and the higher-dimensional model which generalizes the n+1-dimensional
Nambu string model is not known. Even in the 3+1-dimensional case, the identification
of the vortex equation with the complex sine-Gordon equation is not complete.
The variables of the complex sine-Gordon equation has been given in terms
of the vortex string coordinates, but the expression of the vortex string
coordinates in terms of the complex sine-Gordon variables is not
known. Since exact solutions, e.g. solitons and breathers, have
been constructed only in the context of the complex sine-Gordon
equation, the explicit correspondence to the vortex string coordinates is critical
in obtaining exact solutions of the vortex equation systematically using
the inverse scattering method.

In this letter, we resolve these two problems. We first present an
n+1-dimensional generalization of the vortex equation which reduces to the
Nambu string in the no-coupling limit. We identify the
n+1-dimensional vortex equation with the nonabelian sine-Gordon equation and
prove the integrability by finding the associated linear equations of the
inverse scattering.
In doing so, we obtain an expression of vortex coordinates in terms
of the variables of the nonabelian sine-Gordon system. Using this relation,
we obtain explicitly an n-dimensional Hasimoto-type vortex soliton from the one soliton
solution of the nonabelian sine-Gordon equation.

We begin with a review of the vortex model by Lund and
Regge. The relativistic motion of vortices in a uniform static
field is governed by the equation of motion (in a Lorentz frame
in which $X^0 =\t$):
\be
(\pp_{\t}^{2}-\pp_{\s}^{2})X^{i} + c
\e_{ijk}\pp_{\t}X^{j}\pp_{\s}X^{k}=0,  ~; ~ i =1,2,3
\label{vortex}
\ee
and also by the quadratic constraints:
\be
(\pp_{\t}X^{i})^2 +(\pp_{\s}X^{i})^2=1,
~~(\pp_{\s}X^{i})(\pp_{\t}X^{i})=0.
\label{constr}
\ee
Here, $X^{\m }(\s , \t ); \m =0,1,2,3$ are the vortex coordinates and $\s , ~ \t$
are local coordinates on the string world-sheet.
In the no-coupling limit ($c=0$), this equation describes the
transverse modes of the 4-dimensional Nambu-Goto string in the
orthonormal gauge. The critical step leading to the integration of
of the vortex equation (\ref{vortex}) and (\ref{constr}) was to
interpret the equation as the Gauss-Codazzi integrability condition for
the embedding of a surface, i.e. the embedding of the string
world-sheet projected down to the $X^{0}=\t $ hypersurface into the
3-dimensional Euclidean space, $X^{0} = \t $. The induced metric on
the projected world-sheet is given by
\be
ds^2 = ( \pp_{\s }\vec{X})^2 d\s^2 +
2 ( \pp_{\s }\vec{X} \cdot \pp_{\t }\vec{X} ) d\s d\t +
( \pp_{\t }\vec{X})^2 d\t^2 ,
\ee
or
\be
ds^2 =\cos^{2}\phi d\s^2 + \sin^{2}\phi d\t^2
\ee
by parameterizing $( \pp_{\s }\vec{X})^2 =\cos^{2}\phi ,~
( \pp_{\tau }\vec{X})^2 =\sin^{2}\phi $ according
to Eq. (\ref{constr}) . The unit tangent vectors, $\vec{N}_{1}$ and $
\vec{N}_{2}$, spanning the plane tangent to the surface, and the
unit normal vector $\vec{N}_{3}$ consisting a moving frame are given by
\be
\vec{N}_{1} = {1 \over |\pp_{\s}\vec{X}|}\pp_{\s}\vec{X}
,
~~~
\vec{N}_{2} = {1 \over |\pp_{\t}\vec{X}| }\pp_{\t}\vec{X},
~~~
\vec{N}_{3}={1 \over | \pp_{\s }\vec{X} \times \pp_{\t }\vec{X} | }
 \pp_{\s }\vec{X} \times \pp_{\t }\vec{X} .
\ee
The vectors $(\vec{N}_{i}; ~ i=1,2,3)$, given coordinates $u_{1}, u_{2}$
on the surface, satisfy the equation of Gauss and Weingarten:
\be
{\pp \vec{N}_{i} \over \pp u_{k}} = \G^{l}_{ik}\vec{N}_{l} +
L_{ik}\vec{N}_{3}, ~~~
{\pp \vec{N}_{3} \over \pp u_{k}} = -g^{ij}L_{kj}\vec{N}_{i} ,
\ee
where $\G^{l}_{ik}$ are the Christoffel symbols and the $L_{ij}$
are the components of the extrinsic curvature tensor. They are
a set of overdetermined linear equations and the consistency of which requires
the Gauss-Codazzi equation:
\be
R_{ijkl}=L_{ik}L_{jl}-L_{il}L_{jk} , ~~~ L_{ij;k}=L_{ik;j},
\label{Gauss}
\ee
where the semicolon denotes covariant differentiation on the
surface and $R_{ijkl}$ are the components of its Riemann tensor.
From Eq. (\ref{Gauss}), it follows that there exists a field $\eta $
such that
\be
L_{12}=\cot \phi {\pp \eta \over \pp u_{2}} , ~~~
{1 \over 2}(L_{11} + L_{22}) = \cot \phi {\pp \eta \over \pp u_{1}} .
\ee
We introduce the light-cone coordinates $z=(\s + \t )/2 , ~~ \zb =(\s -\t )/2 $
and make the coordinate transformation: $z \rightarrow z/\l , ~~ \zb \rightarrow \l
\zb$ under which the Gauss-Codazzi equation is invariant due to
its Lorentz invariance.
In this case, the Gauss-Weingarten equation in the spin-1/2 representation
changes into the linear equation of the inverse scattering \cite{lund2}:
\be
\pp \Phi = -(U_{0} + \l U_{1})\Phi , ~~~
\pb \Phi = -(V_{0} + \l^{-1} V_{1})\Phi ,
\label{Lax}
\ee
where
\ben
U_{0}+\l U_{1} &=& - \pmatrix{ i c  \l /4 + i \pp \eta \cos 2\phi   / 2 \sin^2 \phi
 & - \pp \phi + i \pp \eta \cot \phi  \cr
\pp \phi + i \pp \eta \cot \phi & -i c \l /4 -i  \pp \eta\cos 2\phi  / 2\sin^2 \phi }
\nonumber \\
V_{0}+\l^{-1} V_{-1} &=& -{i \over 4}\pmatrix{ -c \cos 2\phi /\l - 2 \pb \eta /\sin^2 \phi
&  -c  \sin 2\phi / \l  \cr
-c  \sin 2\phi /\l & c \cos 2\phi /\l + 2 \pb \eta /\sin^2 \phi }.
\label{lin}
\een
The integrability equation:
\be
[\pp +U_{0}+\l U_{1},~ ~  \pb +V_{0}+\l^{-1} V_{-1} ]=0,
\label{int}
\ee
then becomes the complex sine-Gordon equation:
\ben
\pp \pb \phi - {c^2 \over 2} \sin 2\phi + {\cos \phi \over \sin^3
\phi }\pp \eta \pb \eta &=& 0 \nonumber \\
\pb (\cot^2 \phi ~\pp \eta ) + \pp (\cot^2 \phi ~ \pb \eta ) &=& 0.
\label{csg}
\een
This reduces to the well-known sine-Gordon equation when $\eta =0$.
\vglue .2in
Even though the vortex equation as in Eqs. (\ref{vortex}) and (\ref{constr})
has been identified with the complex sine-Gordon equation as in Eq. (\ref{csg}),
the explicit correspondence between variables of each
equations is not well understood. In particular, it is not known
how to write the vortex coordinates $X^{i}$ from the variables of
the complex sine-Gordon system. In order to resolve this problem,
and also to extend the vortex equation to the higher-dimensional case,
we first consider the linear equation in Eq. (\ref{Lax}) and assume that
matrices $U_{0}, U_{1} $ and $ V_{0},  V_{-1}$ are valued in a certain
Lie algebra ${\bf g}$ but otherwise arbitrary. Define
\be
F \equiv \Phi^{-1}(z, \zb , \l ) \l {\pp \over \pp \l } \Phi (z,
\zb , \l ).
\label{vcoor}
\ee
Then, using Eq. (\ref{Lax}), we have
\be
\pp F =-\l \Phi^{-1}U_{1} \Phi , ~~~ \pb F =
{1 \over \l }\Phi^{-1} V_{-1} \Phi .
\label{constr2}
\ee
Also, using Eqs. (\ref{Lax}) and (\ref{int}), we obtain
\be
\pp \pb F = \Phi^{-1}[U_{1}, ~ V_{-1}]\Phi = [\pb F, ~ \pp F] .
\label{gvor1}
\ee
Let $F =\sum \a X^{i}T^{i}~( i =1,...,n=dim {\bf g})$ where
$T^{i}$ are generators of the Lie algebra ${\bf g}$ normalized by
$\mbox{Tr }T^{i}T^{j}=2 \d_{ij}$ and $\a$ is
some constant. Then, Eq. (\ref{gvor1}) becomes
\be
\pp \pb X^{i} = \a f^{ijk} \pb X^{j} \pp X^{k} ,
\label{gvortex}
\ee
where $f^{ijk}$ are structure constants of the Lie algebra ${\bf g}$.
Note that for ${\bf g} \approx so(3)$ this becomes precisely the
vortex equation in Eq. (\ref{vortex}). The constraints as in Eq.
(\ref{constr}), after the coordinate transformation $z \rightarrow z/\l ,
 ~~ \zb \rightarrow \l \zb$, are equivalent to the condition:
\be
\mbox{Tr } (\pp F)^2 = \l^2 \mbox{Tr }(U_{1})^2 = 2 \l^2 \a^2, ~~
\mbox{Tr } (\pb F)^2 ={1 \over \l^2 }\mbox{Tr }(V_{-1})^2 = {1 \over \l^2} 2 \a^2.
\label{gconstr}
\ee
Thus, we define the n-dimensional generalization of the vortex equation in
terms of Eqs. (\ref{gvor1})-(\ref{gconstr}). This equation
is integrable in the sense that it arises from the linear equation
(\ref{Lax}) of the inverse scattering.
In order to better understand the model defined by Eq. (\ref{Lax}) and the
constraint in Eq. (\ref{gconstr}), we first solve the constraint by
fixing $U_{1}$ and $V_{-1}$. By a gauge transformation, we can
always set $U_{1} =T$ for some constant element $T \in {\bf g}$ satisfying
$\mbox{Tr }T^2 = 2  \a^2$.
The remaining constraint, $\mbox{Tr }(V_{-1})^2 = 2 \a^2$, may be
solved for $V_{-1} = g^{-1} \Tb g$ for some constant element $\Tb $
satisfying $\mbox{Tr }\Tb^2  = 2 \a^2$
and an arbitrary group variable $g(z, \zb )$. The zero curvature
condition in Eq. (\ref{int}) should hold for any $\l $, that is,
each coefficients of the polynomial in $\l$ should vanish. Thus,
the coefficients of the $\l $ and the $\l ^{-1}$ terms give rise to
respectively
\be
[\pp + U_{0} , ~ g^{-1}\Tb g ] =0 ~~\mbox{and}~~ [\pb + V_{0} , ~ T] =0,
\ee
which we solve for $U_{0}=g^{-1}\pp g + g^{-1}Ag $ and $V_{0}=\bar{A}$
for some fields $A$ and $\bar{A}$ satisfying the relation
$[A, ~ \Tb ]=0$ and $[\bar{A} , ~ T]=0$. Finally, the zeroth-order
term results in
\be
[\pp + g^{-1}\pp g + g^{-1}Ag , ~ \pb + \bar{A} ] + [T, ~
g^{-1}\Tb g]=0.
\label{nsg}
\ee
This is precisely the nonabelian sine-Gordon equation introduced in the Ref.
\cite{Park1}. We emphasize that, at this stage, $A$ and $\bar{A}$ are regarded only
as background fields which commute with arbitrary constant elements $\Tb$ and $T$
respectively. Further specifications of these variables and their physical meanings
are given below. As for the field theory formulation,
one can readily check that the nonabelian sine-Gordon
equation (\ref{nsg}) arises from the gauged Wess-Zumino-Novikov-Witten action plus
a potential term:
\ben
S &=& S_{WZNW}(g) + S_{\mbox{gauge}} - S_{\mbox{pot}}
\label{action} \\
S_{WZNW}(g)&=& -{1\over 4\pi }\int_{\S }dz d\zb
\mbox{Tr }( g^{-1} \pp g g^{-1} \pb g) - {1 \over 12\pi }\int_{B}\mbox{Tr }
(\tilde{g}^{-1}d \tilde{g}\wedge \tilde{g}^{-1}d \tilde{g} \wedge
\tilde{g}^{-1}d \tilde{g}) \ , \nonumber \\
\label{wzw}
S_{\mbox{gauge}} &=&
{1 \over 2\pi }\int \mbox {Tr} (- A\pb g g^{-1} + \bar{A} g^{-1} \pp g
 + Ag\bar{A} g^{-1} - A\bar{A} ) \ , \nonumber \\
S_{\mbox{pot}}&=& {1 \over 2\pi }\int dz d\zb \mbox{Tr}(g\Tb g^{-1} T) ,
\nonumber
\een
where $S_{WZNW}(g)$ is the usual Wess-Zumino-Novikov-Witten action.

The nonabelian sine-Gordon model in Eq. (\ref{nsg}) associated
with a Lie algebra ${\bf g}$ is rather general for the
practical purpose of obtaining exact solutions. Thus, we make further
restrictions by specifying $T$ and $\Tb$ as follows;  we
assume that $T $ and $ \Tb$ belong to the Cartan subalgebra of ${\bf g}$
and the subalgebra ${\bf h} \subset {\bf g}$ is the common
centralizer of $T$ and $\Tb $, i.e. $ {\bf h} = \{ h \subset {\bf g}: [T,~ h]
=0, ~[ \Tb, h]=0 \}$. We also assume $A$ and $\bar{A}$ to be
valued in ${\bf h}$ so that the gauged Wess-Zumino-Novikov-Witten
action, $S_{WZNW}(g) + S_{\mbox{gauge}}$, becomes the $G/H$-WZNW action
for the coset conformal field theory \cite{coset}. Note that the potential
term $S_{\mbox{pot}}$ is invariant under the $H$-group action
so that the whole action $S$ possesses the group $H$-vector gauge
invariance if we treat $A$ and $\bar{A}$ as gauge connections.
Moreover, we could further restrict the model by treating $A$ and $\bar{A}$
as Lagrangian multipliers which give rise to the constraint equations,
\be
( \ - \pb g g^{-1} + g\bar{A} g^{-1} - \bar{A} \  )_{\bf h} = 0 , ~~
( \  g^{-1} \pp g  +g^{-1} A g - A )_{\bf h} = 0 .
\label{const3}
\ee
Here, the subscript ${\bf h}$ denotes the projection to the
subalgebra ${\bf h}$. This restricted nonabelian sine-Gordon model
corresponding to the coset $G/H$ has been named as the symmetric
space sine-Gordon(SSSG) model for the type-II symmetric spaces
\cite{bps}. It has been also shown that the field strength of
$A, \bar{A}$ vanishes, i.e. $F_{z\zb }=[\pp + A, ~ \pb + \bar{A}]$.
Thus, we may fix the vector gauge invariance by taking
$A=\bar{A}=0$. Note that the vortex string coordinates in Eq.
(\ref{vcoor}) are also invariant under the vector gauge
transformation: $ \Phi \rightarrow h\Phi $ for an element $h(z, \zb) \in
H$. This means that the vortex solution $X^{i}$ can be obtained from the solution
of the gauge fixed $(A=\bar{A}=0)$ SSSG equation:
\be
-\pb( g^{-1}\pp g ) + [T, ~ g^{-1}\Tb g]=0 ,
\label{nsg1}
\ee
and the constraints,
\be
(g^{-1} \pp g)_{\bf{h}}=0, ~~~ (\pb g g^{-1})_{\bf{h}}=0.
\ee
Other types of symmetries of the vortex and the SSSG models are also
interconnected. For example, the symmetry of the SSSG equation under the parity
transformation: $g \rightarrow P g,\ \ \zb \rightarrow - \zb$, for $P$ a
constant element which anti-commutes with $T$ and $ \Tb$, induces a
symmetry of the vortex equation under the exchange( $\t \leftrightarrow
\s$) of string world-sheet coordinates. On the other hand, the
transformation: $\Phi(\l, z, \zb) \rightarrow \Phi(\l, z, \zb) \tilde{\Phi}(\l)$
for a unitary element $\tilde{\Phi}(\l)$, which leaves the linear
equation (\ref{Lax}) invariant, induces the rotational and the translational
transformation a vortex such that
\be
F \rightarrow \tilde{\Phi}(\l)^{-1} F \tilde{\Phi}(\l) + \l \tilde{\Phi}(\l)^{-1}
{d \tilde{\Phi}(\l) \over d \l}.
\label{symme}
\ee
If $\tilde{\Phi}(\l)$ is not unitary, e.g. $\tilde{\Phi}(\l)=f(\l )$
for some function $f(\l )$, the trace of $F$ changes.
Thus we can always set the trace to zero by choosing an appropriate $f(\l )$.

Next, we derive a vortex solution from the one soliton solution
of the SSSG equation. Instead of applying the method of inverse
scattering, we adopt the following B\"{a}cklund transformation to
derive the one soliton solution; let $(f, \Phi_{f}) $ is a solution of the
linear equation (\ref{Lax}). Then, $(g, \Phi_{g})$ is another solution
provided that
\be
\Phi_{g} = {\l \over \l -i b }\Big( 1 + {i b \over \l }g^{-1} f \Big)
\Phi_{f}
\label{BT1}
\ee
and
\ben
g^{-1} \pp g - f^{-1} \pp f - ib [g^{-1} f, ~ T] &=& 0 \nonumber \\
ib \pb (g^{-1} f) + g^{-1} \Tb g - f^{-1} \Tb f &=& 0,
\label{BT2}
\een
where $b$ is a parameter of the B\"{a}cklund transformation.
For simplicity,  we take a trivial solution for $(f, ~ \Phi_{f})$ such
that
\be
f=1, ~~~ \Phi_{f} = exp(-\l T z -\l^{-1}\Tb \zb ) .
\label{triv}
\ee
One can easily see that this corresponds to the straight vortex line
$F_{f}=-\l T z + \l^{-1} \Tb \zb$. In order to solve Eq.
(\ref{BT2}) with the trivial solution in Eq. (\ref{triv}),
we use the fact that $g^{-1}\pp g$ is
anti-Hermitian so that $[g - g^{-1}, ~T] =0$ due to Eq.
(\ref{BT2}). This may be solved in terms of a Hermitian projection matrix
 $P$ satisfying $ ~ P^2 = P, ~ P^{\dagger }=P$ by
\be
g=2 \cos \theta P - e^{i \theta} .
\ee
where $\theta$ is some constant parameter. The linear equation now
changes into
\be
(1-P)(\pp - i be^{i \theta}T)P=0 , ~~~ (1-P)(ibe^{i \theta} \pb-\Tb )P=0 .
\label{plin}
\ee
If we consider only the 1-dimensional projection, we may write
\be
P_{ij}=s_{i}s_{j}^{*} / \sum_{k=1}^{n}s_{k}s_{k}^{*}
\label{P1}
\ee
so that Eq. (\ref{plin}) becomes
\be
(\pp - i be^{i \theta}T)\vec{s}=0, ~~~ (ibe^{i \theta} \pb -\Tb )\vec{s}=0 .
\ee
This can be integrated immediately to yield
\be
s_{i}=\sum_{k=1}^{n}[exp(i be^{i \theta}Tz-i e^{-i \theta}\Tb \zb
/b)]_{ik}u_{k},
\label{P2}
\ee
where $u_{i}$ are constants of integration. Finally, using
\be
\Phi_{g}=f(\l ) \Big( 1 + {i b \over \l }
(2 \cos \theta P -e^{-i \theta})\Big)exp(-\l T z -\l^{-1}\Tb \zb )
\ee
where $f(\l )$ is chosen to make $F$ traceless,
we obtain the n-dimensional vortex soliton solution of the
Hasimoto-type \cite{hasimoto}:
\be
F= {2 i b\l  \cos \theta \over 2b \l \sin \theta -\l^2 -b^2 }\Big[ exp(\l T
z + \l^{-1}\Tb \zb )Pexp(-\l T z - \l^{-1}\Tb \zb )-1/2 \Big]  - \l T z +
\l^{-1} \Tb \zb ,
\label{Has}
\ee
where $P$ is defined by Eqs. (\ref{P1}) and (\ref{P2}).

Now,  we restrict to the case of Lund and Regge by taking ${\bf g} \approx
so(3)$ and $T^i = \s _i$ where $\s _i$ are Pauli matrices. $c$ and $\a$ as in Eqs.
(\ref{vortex}) and (\ref{gvortex}) are related by $c = -4 i \a$.
Choosing $T$ and $\Tb$ by $T=-\Tb=-i  c \s_{3}/4$
and also with an appropriate parametrization of an $SU(2)$ element $g$,
one can readily see that the SSSG equation becomes the complex
sine-Gordon equation in Eq. (\ref{csg}).
The vortex coordinates $X^{i}$ in Eq. (\ref{vortex}) are given by
the components of $F$ with the following scaling;
\be
X_i = -{4 i \over c} F_i\Big( z={1 \over 2 \l}(\s + \t), \zb ={\l \over 2} (\s - \t)
\Big) .
\label{realco}
\ee
Then, the vortex soliton in Eq. (\ref{Has}) becomes
\ben
X_1 &=& R \rm{sech} \S \ \cos \T, \nonumber \\
X_2 &=& R \rm{sech} \S \ \sin \T, \nonumber \\
X_3 &=& R \tanh \S + \s,
\een
where
\ben
R &=& {4 b \l \cos \q \over c (2b \l \sin \q -\l^2 -b^2)}, \nonumber \\
\S &=& {c \over 4} \cos \q \left( {b \over \l} (\s +\t)
+ {\l \over b} (\s- \t) \right), \nonumber \\
\T &=& {c \over 2} \t +{c \over 4} \sin \q \left( {\l \over b}
( \s -\t)-{b \over \l} ( \s + \t) \right).
\een
In this paper, we have extended the vortex equation by Lund and
Regge to the n-dimensional case and proved its integrability by
mapping the vortex equation into the nonabelian sine-Gordon equation
defined in association with a Lie algebra ${\bf g}$ of dimension
$n$. Through the identification, we have obtained explicit
correspondence between vortex coordinates and the variables of the
nonabelian sine-Gordon system, and also the Hasimoto-type one soliton
solution for the vortex equation. Other explicit solutions can be
also found through this correspondence with interesting physical
implications. This will appear elsewhere \cite{Lee}.

\vglue .2in
{\bf ACKNOWLEDGEMENT}
\vglue .2in
We are grateful to K. Lee for many helpful discussions.
This work was supported in part by the program of Basic Science Research,
Ministry of Education 1998-015-D00073, and by Korea Science and Engineering
Foundation, 97-07-02-02-01-3.

\end{document}